\documentclass[aps,pre,reprint,showpacs,bibnotes,twoside,eqsecnum]{revtex4-1}

\usepackage{amsmath}
\usepackage{amssymb}
\usepackage{epsfig}
\usepackage{bm}
\usepackage{latexsym}
\usepackage{color}
\usepackage{mathtools}
\usepackage{epstopdf}

\begin{document}

\title{Non-standard anomalous heat conduction in harmonic chains with correlated isotopic disorder}

\author{I.~F.~Herrera-Gonz\'alez}
\affiliation{Department of Engineering, Universidad Popular Aut\'onoma del Estado de Puebla, 21 Sur 1103, Barrio Santiago, Puebla, Pue., M\'exico}
\author{J.~A.~M\'endez-Berm\'udez }
\affiliation{
Departamento de Matem\'{a}tica Aplicada e Estat\'{i}stica, Instituto de Ci\^{e}ncias Matem\'{a}ticas e de Computa\c{c}\~{a}o, Universidade de S\~{a}o Paulo - Campus de S\~{a}o Carlos, Caixa Postal 668, 13560-970 S\~{a}o Carlos, SP, Brazil \\
and \\
Instituto de F\'isica, Benem\'erita Universidad Aut\'onoma de Puebla, Apartado postal J-48, Puebla 72570, M\'exico}

\date{\today}

\begin{abstract}
We address the general problem of heat conduction in one dimensional harmonic chain, with correlated isotopic disorder, attached at its ends to white noise or oscillator heat baths. When the low wavelength $\mu$ behavior of the power spectrum $W$ (of the fluctuations of the random masses around their common mean value) scales as $W(\mu)\sim \mu^\beta$, the asymptotic thermal conductivity $\kappa$ scales with the system size $N$ as $\kappa \sim N^{(1+\beta)/(2+\beta)}$ for free boundary conditions, whereas for fixed boundary conditions $\kappa \sim N^{(\beta-1)/(2+\beta)}$; where $\beta>-1$, which is the usual power law scaling for one dimensional systems. Nevertheless, if $W$ does not scale as a power law in the low wavelength limit, the thermal conductivity may not scale in its usual form $\kappa\sim N^{\alpha}$, where the value of $\alpha$ depends on the particular one dimensional model.  As an example of the latter statement, if $W(\mu)\sim \exp(-1/\mu)/\mu^2$, $\kappa \sim N/(\log N)^3$ for fixed boundary conditions and $\kappa \sim N/\log(N)$ for free boundary conditions, which represent non-standard scalings of the thermal conductivity. 
\end{abstract}


\maketitle

\section{Introduction}

Understanding the statistical properties of open, many-particles systems is one of the challenges of nonequilibrium statistical mechanics. 
In particular, deriving the phenomenological laws from the microscopic dynamics is an actual unsolved problem. For instance, one of such phenomenological equations is Fourier's law. Moreover, one dimensional (1D) harmonic chains are suitable models to study heat conduction due to the fact that they are analytically tractable and nonlinear effects can be neglected in the low-temperature regime. Thus, heat conduction in harmonic chains has been intensively studied in the last decades. One of the main results is that thermal conductivity scales with the system size as 
\begin{equation}
\kappa\sim N^{\alpha},
\label{scale} 
\end{equation}
where the value of $\alpha$ depends on the particular 1D model: For an ordered harmonic chain, attached at its ends to Langevin heat baths, $\alpha=1$ regardless on the boundary conditions~\cite{CL71}; whereas when uncorrelated disorder is introduced, $\alpha=-1/2$ for free boundary conditions and $\alpha=1/2$ for fixed boundary conditions~\cite{MI70,I73}. The value of $\alpha$ depends on the thermal properties of the heat bath as well, a suitable choice of the latter can even lead to finite thermal conductivity~\cite{D01}. Although, when the uncorrelated disordered harmonic chain is attached at its ends to oscillator baths (Rubin and Greer's heat baths) the thermal conductivity scales as in the Langevin heat bath case~\cite{RG71,V79,DD08}. The role that correlations of the random masses plays in the heat conduction has been studied as well, where analytical results for the asymptotic thermal conductivity showed that the $\alpha$ parameter can be controlled through the low wave number form of the power spectrum~\cite{HM19,ZLW15,HIT15}. Analytical results for the $\alpha$ exponent were also obtained when the springs of the harmonic chain are randomly distributed either with box or power law distributions~\cite{AOYI18,AABOI19}, where the results are still valid in the quantum regime.

The size dependence of the thermal conductivity persists, for the vast majority of the 1D models, even when one introduces phonon-phonon scattering to account for nonlinearity in the system~\cite{LLP03,D08,LXXZZL12}. Thus, Fourier's law is not satisfied for most 1D models since the thermal conductivity should be an intensive quantity. If the thermal conductivity depends on the system size the heat conduction is known as {\it anomalous heat conduction}, and if this size-dependence is particularly given by Eq.~(\ref{scale}) we label it as {\it standard-anomalous heat conduction}. The reason of this designation comes from the fact that, for all the 1D systems that have been analyzed, the anomalous heat conduction scales as Eq.~(\ref{scale}). In contrast, for two dimensional non-linear models the heat conduction is non-standard since $\kappa\sim \log N$, whereas in three dimensions $\kappa$ becomes an intensive quantity  as predicted by the mode-coupling theory~\cite{LLP03}. The anomalous heat conduction behavior predicted by some of these low dimensional lattices has been observed in carbon and boron-nitride nanotubes~\cite{COGMZ08} and in other quasi 1D nanostructures; see e.g.~Ref.~\cite{LXXZZL12}.

Now, the following question arises: Is it possible to find a 1D system displaying a non-standard anomalous heat conduction? The answer of this question was partially given in Refs.~\cite{HM19,HIT15}. In Ref.~\cite{HM19} the heat conduction in harmonic chains with L\'evy--type disorder was studied; there it was shown that the asymptotic thermal conductivity acquires a non-closed form when the L\'evy parameter equals 2. In Ref.~\cite{HIT15},  analytical estimations showed that using specific long--range correlations of the disorder, the asymptotic thermal conductivity results to be non-standard, but this asymptotic behavior is only reached for huge system sizes ($\ln N \gg 1$). Thus numerical confirmation was not possible in either case. In addition, all the results derived in both works were restricted to the case of weak coupling between the harmonic chain and the Langevin heat baths. 

The aim of the present work is twofold: On the one hand, we propose a specific correlated disorder model where non-standard anomalous heat conduction can be observed numerically. On the other hand, we extend the range of validity of the results given in Ref.~\cite{HIT15}, since we show that they are valid for any coupling between the harmonic chain and Langevin or oscillator heat baths.

Below, we will refer to Langevin heat baths and oscillator heat baths (formed of harmonic oscillators of equal masses $m$ and springs constant $k$) as Model I and Model II, respectively.

\section{The model}

The Hamiltonian of the isotopically disordered harmonic chain is given by 
\begin{equation}
H=\sum^N_{n=1} \frac{p^2_n}{2m_n} +\frac{1}{2}k\sum^{N-1}_{n=1}(x_{n+1}-x_n)^2+\frac{1}{2}k_1(x^2_1+x^2_N),
\label{Eq2}
\end{equation}
where ($x_n,p_n$) denote the displacement and the momentum of the $n$-th particle of the lattice. The masses $m_n$ are correlated random variables whose statistical properties will be specified later. On-site potentials of strength $k_1$ have been added  at the boundaries of the chain to reproduce different boundary conditions as we will define below. 

When the system is attached to heat baths, the motion of the first and last particles are affected directly by the retarded dissipation $\Sigma_{L,R}$ and noise $\eta_{L,R}$ terms provided by the left ($L$) and right ($R$) reservoirs. Thus, the equations of motion are given by the following generalized Langevin equations~\cite{DD08}
\begin{eqnarray}
m_N\ddot{x}_N&=&kx_{N-1}-Bkx_{N}-k_0x_N+ \Sigma_R \ast x_N+\eta_R(t) , \nonumber \\  
m_n\ddot{x}_n&=& k\left(x_{n-1}+x_{n+1}-2x_n\right) , \quad n=2\ldots N-1 , \nonumber \\
m_1\ddot{x}_1&=&kx_2-Bkx_1-k_0x_1+ \Sigma_L\ast x_1+\eta_L(t) .
\label{Langevin}
\end{eqnarray}
Here $(\Sigma\ast x)(t)=\int^t_{-\infty}\Sigma(t-t')x(t')dt'$ denotes the convolution integral of $\Sigma$ and $x$, $k_0=k_1$ ($k_0=k_1-k$, $k_1>k$) for Model I (Model II), and $B=1$ ($B=2$) for Model I (Model II). The parameter $B$ takes the value of 2 for Model II because the term $kx_1$ ($kx_N$) is incorporated into the equations of motion to account for the interaction of the oscillator baths with the first (last) particle of the harmonic chain, see details in Ref.~\cite{OL74}. Therefore, free boundary conditions are determined by setting $k_1=0$ for Model I and $k_1=k$ for Model II; whereas $k_1\ne0$ and $k_1\ne k$ correspond to fixed boundary conditions for Model I and Model II, respectively. Thus, for fixed boundary conditions, the outermost masses of the chain are coupled to external ``walls" of infinite mass with spring constants $k_0$.
 
The noise $\eta_{R,L}$ and the retarded dissipation $\Sigma_{L,R}$ terms are not independent quantities, they must satisfy the {\it fluctuation dissipation theorem} (FDT) in order to ensure thermal equilibrium in the heat baths. Thus, FDT can be cast in the following useful form~\cite{K78}
\begin{eqnarray}
\langle \hat {\eta}_{L,R}(\omega)\hat {\eta}_{L,R}(\omega')\rangle&=&  S_{\eta_{L,R}}(\omega)\delta(\omega+\omega'), \nonumber \\
S_{\eta_{L,R}}(\omega)&=&-\frac{k_BT_{L,R}\text{Im}[\hat{\Sigma}_{L,R}(\omega)]}{\pi\omega}.
\label{disi}
\end{eqnarray}
Here, $\hat {\eta}_{L,R}(\omega)$ is the Fourier transform of $\eta_{L,R}(t)$, the angular brackets $\langle \cdot \rangle$ represent noise average, $S_{\eta_{L,R}}(\omega)$ denotes the power spectrum of the noises $\eta_{L,R}(t)$,
\begin{equation} 
\Gamma(\omega)\equiv\text{Im}[\hat{\Sigma}_{L,R}(\omega)]=\text{Im}\left[\int^{\infty}_0 \Sigma_{L,R}(t)e^{i \omega t}\right]
\label{g}
\end{equation}
is the imaginary part of the Fourier-Laplace transform of $\Sigma_{L,R}(t)$, and $k_B$ is the Boltzmann constant.  Therefore, the spectral properties of the heat bath are completely determined by $\text{Im}[\hat{\Sigma}_{L,R}(\omega)]$.

For Model I, it is easy to prove that
\begin{eqnarray} 
\hat{\Sigma}(\omega)=-i\gamma\omega, \qquad \Gamma(\omega)=-\gamma\omega ,
\label{g1}
\end{eqnarray}
with $\gamma$ representing the coupling strength of the baths with the harmonic chain. While, for Rubin-Greer's baths \cite{OL74}
\begin{eqnarray}
\hat{\Sigma}(\omega)&=&k \left\lbrace1-m\frac{\omega^2}{2k}-i\omega\left(\frac{m}{k}\right)^{1/2}\left[1-\frac{\omega^2}{4k} \right]^{1/2}\right\rbrace , \nonumber \\
\Gamma(\omega)&=&-k\omega\left(\frac{m}{k}\right)^{1/2}\left[1-\frac{\omega^2}{4k} \right]^{1/2} .
\label{g2}
\end{eqnarray}

\section{Heat conduction in the harmonic chain}
\label{heat}

The steady state classical heat current (heat flux) through the harmonic chain can be calculated using the following formula \cite{DD08}
\begin{eqnarray}
J_N=\frac{(T_L-T_R)}{4\pi}\int^{\infty}_{-\infty}d\omega 4\Gamma^2(\omega) \left |\hat{G}_{1N} \right|^2
\label{flux}
\end{eqnarray}
with
\begin{eqnarray*}
\left|\hat{G}_{1N} \right|^2=\left|D_{1N}-\Sigma'\left(D_{2N}-D_{1N-1} \right)+\Sigma'^2D_{2N-1} \right|^{-2},
\end{eqnarray*}
here the Boltzmann constant has been set to one and $\Gamma(\omega)$ is determined from Eqs.~(\ref{g1}) or~(\ref{g2}), depending on the chosen model of thermal baths. The matrix elements $D_{lm}$  depends only on the characteristics of closed harmonic chain and are given by the following product of random matrices 
\begin{eqnarray}
        \left(
                \begin{array}{rr}
                 D_{1N} & -D_{1N-1} \\
                 D_{2N} &  -D_{2N-1} 
                \end{array}
         \right)=T_1T_2 \dots T_N,
\label{matrix1}
\end{eqnarray}
where
\begin{eqnarray}
       T_n =\left(
                \begin{array}{rr}
                  2-m_n\omega^2/k & \ \ -1 \\
                  1 \ \ \ \ \ \ \ &  0. 
                \end{array}
         \right).
\label{Tr}
\end{eqnarray}

The information of the heat baths and boundary conditions are contained in the term $\Sigma'=\hat{\Sigma}/k-k_1/k+1$. In addition,  the integrand of Eq.~(\ref{flux}) can be identified as the transmission coefficient of phonons through the disordered chain given by 
\begin{eqnarray}
{\cal T}_N=4\Gamma^2(\omega) \left |\hat{G}_{1N} \right|^2.
\label{trans} 
\end{eqnarray}

By taking into account Fustemberg theorem~\cite{F63}, it is possible to probe rigorously that  the elements of the matrix (\ref{matrix1}) grows exponentially with $N$ (if $N$ is large enough) at the rate~\cite{I73}
\begin{equation}
\lambda(\omega)=\lim_{N\rightarrow \infty} \frac{1}{N}\left|T_1T_2\dots T_N {\bf u}\right|,
\label{cr}
\end{equation}
with ${\bf u}$ any non-zero vector of two components, and $T_i$ are {\it uncorrelated random matrices}. For the case of matrices with correlated disorder there is no general prove for the exponential growth of the elements of the matrix~(\ref{matrix1}). In fact, there are some correlated disorder models for which $\lambda(\omega)=0$ for $\omega \ne 0$ like the random dimer model~\cite{BG20}, whereas for other models the rate of exponential growth only vanishes within the second order of a perturbative approach, in the case of weak disorder~\cite{IKM12,HIT10,L02}. Thus, we only focus our study on correlated disordered models for which $\lambda(\omega)>0$ for $\omega \ne 0$
 
Since the rate of exponential growth is gauged by the frequency $\omega$, which can be  seen in the structure of the matrix $T_n$, for large enough chains, only the lowest frequency components of the matrix (\ref{matrix1}) will contribute to the heat flux. Therefore, one can introduce the concept of cut-off frequency  $\omega_c$ which determines how many matrix components $\hat{G}_{1N}$  must be taken into account to compute the heat flux; matrix components $\hat{G}_{1N}$ corresponding to frequencies below $\omega_c$ will contribute to the heat flux, in contrast, their contribution to the heat transport can be neglected for frequencies which are above $\omega_c$, when $N\gg1$. It is important to notice that $\omega_c(N)$ depends on  the system size and the statistical properties of the random masses; also $\omega_c\rightarrow 0$ if $N\rightarrow \infty$. 

If the length $N$ of the system is smaller than the mean free path $l_f(\omega)$, the probability of phonon scattering is practically negligible and the transmission coefficient of the disordered harmonic chain ${\cal T}_N$ is practically equal to the transmission coefficient of the corresponding ordered chain ${\cal T}^O_N$  (${\cal T}_N \approx {\cal T}^O_N$). If $l_f\lesssim N$, the exponential growth of the elements of the matrix (\ref{matrix1}) arises. Therefore, using the above arguments, we can write 
\begin{eqnarray}
 J_N \sim (T_L-T_R)\int^{\omega_c}_0{\cal T}^O_N(\omega)d\omega,
\label{flux1}
\end{eqnarray} 
where disorder fluctuations of $J_N$ go to zero in the thermodynamic limit due to the fact that $\lambda(\omega)$ is a self-averaged quantity.
 
One can obtain the transmission coefficient for an ordered chain by diagonalizing the matrices given in Eq. (\ref{Tr}), then the product of random matrices~(\ref{matrix1}) is easily computed, and after replacing these  results in Eq. (\ref{trans}), we get
\begin{eqnarray}
    {\cal T}^O_N=
         \left\{
                \begin{array}{ll}
                {\displaystyle
                  \frac{\gamma \omega^2\sqrt{4Mk-M\omega^2}}{k^2_1+\left[\gamma^2 -Mk_0\right]\omega^2} } & \quad \mbox{for Model I},  \vspace{0.2cm} \\     
                  {\displaystyle            
                   \frac{2kM\omega^2\left(1-M\omega^{2}/4k \right)}{2kM\omega^2\left(1-M\omega^{2}/4k \right)+k_0} } & \quad \mbox{for Model II} ; 
                \end{array}
         \right. 
\label{nosen}
\end{eqnarray} 
expressions valid for large system sizes $N$. Above, $M$ is the the value of the masses of the ordered harmonic chain, which in the disordered model corresponds to the first moment of the masses, and $\Sigma'$ has been replaced by its low frequency form up to first order in $\omega$ (since $\omega_c$ is a monotonically decreasing function of $N$, $\omega \ll 1$ for large system sizes). Notice that for Model II, and free boundary conditions, Eq.~(\ref{nosen}) predicts ${\cal T}^O_N=1$ as it should be. 

Now, the asymptotic form of $J_N$ can be written in terms of $w_c(N)$ if one uses Eq.~(\ref{nosen}) into expression~(\ref{flux1}) and takes into account that $\omega_c \ll 1$. Therefore, for free boundary conditions one has for both models of heat baths that
\begin{equation}
 J_N  \sim  \omega_c(N),
\label{a1} 
\end{equation}
whereas, for pinned boundary sites, $J_N$ takes the form
\begin{equation}
 J_N  \sim  \omega^3_c(N).
\label{a2} 
\end{equation} 

When one compares the above results with the corresponding ones given in Ref.~\cite{HIT15}, it is possible to conclude that 
\begin{eqnarray}
\omega_c\sim \frac{N_e}{N},
\end{eqnarray}
where $N_e$ represents the number of low frequency extended vibrational modes that contribute to the heat transport. However, this result is only valid for weak coupling between the harmonic chain and the Langevin thermal baths. When a larger coupling is considered, cross-coupling between the dynamics of the different eigenmodes must be considered. Thus, the great advantage of equations~(\ref{a1}) and~(\ref{a2}) is that they are still valid even for strong coupling, and for both models of heat baths.    
 
Now, the key point is to find the cut-off frequency in terms of the system size.

\section{Localization properties of the correlated disorder chain}

When one considers uncorrelated isotopic disorder in the harmonic chain, its vibrational modes become exponentially localized in the thermodynamic limit with the characteristic length $L_{\text{loc}}$ which is known as the localization length~\cite{I73}; this exponential localization persists for the correlated disordered models studied in this work. The only mode which is extended is the zero frequency mode. Indeed, the localization length is given by the following expression~\cite{HM19}   
\begin{eqnarray}
L^{-1}_{\text{loc}}(\mu)=\frac{\mbox{var}[m_n]}{2M^2}\tan^2\left(\frac{\mu}{2}\right)W(\mu),
\label{locEPL}
\end{eqnarray} 
where
\begin{eqnarray}
W(\mu)=1+2\sum^{\infty}_{l=1}\chi(l)\cos(2l\mu)
\label{powern}
\end{eqnarray} 
is the power spectrum of the fluctuations of the random masses around their common mean value $M$, $\mbox{var}[m_n]$ denotes the variances of the disorder masses, $\mu$ is the wave number of the plane waves  which are the solutions of the dynamical equations for an ordered harmonic chain. Thus, $\mu$ is related to the frequency $\omega$ through the dispersion relation  $\omega(\mu)=\omega_{\text{max}}\left|\sin\left(\mu/2\right)\right|$. $\chi(l)=\overline{\delta m_n \delta m_{n+l}}/\text{var}[m_n]$ is the normalized binary correlator where $\overline{x}$ depicts disorder average of $x$.

Equation~(\ref{locEPL}) is valid at the effective weak disorder condition
\begin{equation}
\left(\frac{2\omega}{\omega_{\text{max}}}\right)^2\frac{\sqrt{\mbox{var}[m_n]}}{M}\ll1,
\label{weak}
\end{equation}
where
\begin{equation}
\omega_{\text{max}}=\sqrt{\frac{4k}{M}}
\end{equation}
is the largest frequency of the vibrational modes of an ordered chain. In the low frequency limit Eq.~(\ref{locEPL}) gets the form 
\begin{equation}
L^{-1}_{\text{loc}}(\omega)=\frac{\mbox{var}[m_n]}{2 M^2}\left(\frac{\omega}{\omega_{\text{max}}} \right)^2W\left(\frac{2\omega}{\omega_{\text{max}}}\right), \ \ \text{when} \ \ \omega \rightarrow 0,
\label{locw}
\end{equation}
an equation which has also been derived in Refs.~\cite{ZLW15,HIT15}.

\section{Thermal conductivity of the chain with correlated disorder}

\subsection{Standard thermal conductivity}

Now it is possible to estimate the cut-off frequency $\omega_c$ by considering the following arguments: The principal contribution to the heat flux comes from the low frequency dependence of the transmission coefficient ${\cal T}_N$, which is practically equal to the transmission coefficient of the ordered chain as  explained in Sec.~\ref{heat}. Furthermore, if one takes into account that the localization $L_{\text{loc}}$  is the characteristic length for an exponentially localized vibrational mode which is an extended state if $N\le L_{\text{loc}}(\omega)$, thus by definition, the cut-off frequency provides the equality in the latter equation. Therefore, using Eq.~(\ref{locw}), we get 
\begin{equation}
\omega_c=M\omega_{\text{max}}\sqrt{\frac{2}{\text{var}[m_n]N}\frac{1}{W\left(\frac{2\omega_c}{\omega_{\text{max}}} \right)}} \ ;
\label{apro1}    
\end{equation}
which stablishes the size dependence of $\omega_c$ and, in combination with Eqs.~(\ref{a1}) and~(\ref{a2}), determines the asymptotic scaling law of the heat flux $J_N$ with the system size (that depends on the boundary conditions).

The most common situation is when the power spectrum scales as a power law of $\omega$, in the low frequency domain, 
\begin{eqnarray}
{W}(\omega) \propto \omega^\beta \ \ \mbox{with} \ \ \beta>-1.
\label{low}
\end{eqnarray}
Here, $\beta>-1$ since from Eq.~(\ref{powern}) it is seen that the power spectrum must satisfy the normalization condition 
\begin{eqnarray}
\int^{\frac{\pi}{2}}_0 {W}(\mu)d\mu=\frac{\pi}{2} \ .
\label{nor}
\end{eqnarray}
Thus, if relation (\ref{low}) holds in the low frequency domain then a clear dependence of $\omega_c$ in terms of $N$ is obtained:
\begin{equation}                  
\omega_c \sim \left(\frac{1}{N}\right)^{1/(2+\beta)}. 
\label{aprof}
\end{equation}
For the uncorrelated disorder case, $\beta=0$, the latter scaling law is equal to the one provided by Matsuda and Ishii~\cite{MI70}. 

By replacing the scaling law~(\ref{aprof}) into Eqs.~(\ref{a1}) and~(\ref{a2}) and taking into account that $\kappa= J_N  N/(T_L-T_R)$ we finally get the asymptotic form of $\kappa$, which is the same for Models I and II, and has the following standard form:
\begin{eqnarray}
 \kappa\sim N^{\alpha}, \ \  \alpha=
         \left\{
                \begin{array}{ll}
                {\displaystyle \frac{1+\beta}{2+\beta}  }  & \mbox{for free BC}   \vspace{0.2cm} \\     
                  {\displaystyle \frac{\beta-1}{2+\beta} } & \mbox{for fixed BC} 
                \end{array}
         \right. .
\label{final}
\end{eqnarray} 
When the boundaries of the chains are subject to on-site potentials (fixed BC), $\alpha>-2$ because $\beta>-1$, and the choice $\beta=1$ gives the Fourier's law. For free BC, $\alpha>0$ and the thermal conductivity diverges in the thermodynamic limit. Nevertheless, if $\beta\rightarrow -1$, the system exhibits normal heat conduction asymptotically. When $\beta\rightarrow \infty$, the harmonic system interacting with either models of heat baths tends towards the ballistic regime $\kappa \sim N$.

\begin{figure}[!t]
\centering
\includegraphics[width=\columnwidth]{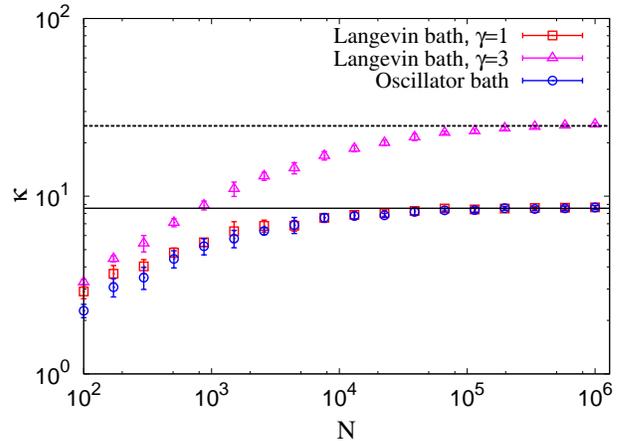}
\caption{(Color online) Thermal conductivity as a function of the chain size for fixed boundary conditions. $T_L-T_R=1$ for both bath types. For oscillator baths $k_2=k_1=k=1$. The black continuous line is the average of the thermal conductivity of the last 4 numerical data, while the value of the dashed line is three times the value of the  continuous line.}
\label{figuFC}
\end{figure}

\begin{figure}[!h]
\centering
\includegraphics[width=\columnwidth]{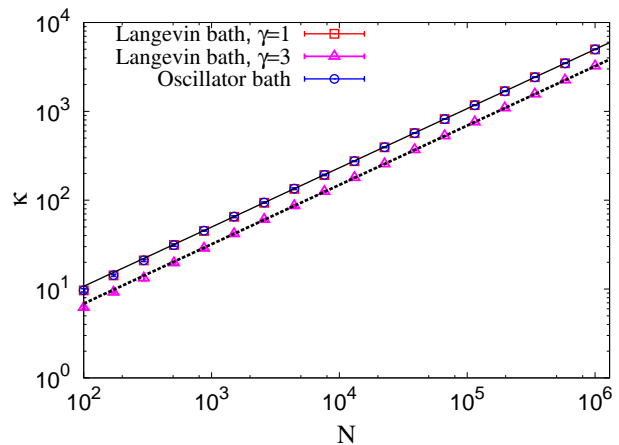}
\caption{(Color online) Thermal conductivity as a function of the chain size for free boundary conditions. $T_L-T_R=1$ for both bath types. For oscillator baths $k_2=k_1=k=1$. The black continuous (dashed) line is the best fit of the numerical data with the function $f(N)=aN^b$ for $\gamma=1$ ($\gamma=3$). For $\gamma=1$ we got $a=0.499 \pm 0.003$ and $b=0.6666\pm 0.0004$, while for $\gamma=3$ we obtained $a= 0.316 \pm 0.001$ and $b= 0.668 \pm 0.003$.}
\label{figuLC}
\end{figure}

In order to verify our analytical result~(\ref{final}), we propose the power spectrum
\begin{equation}
W(\mu) = \frac{\pi (\sqrt{2} + 1)}{2 \sqrt{2}} \sin \left( \frac{\mu}{2} \right) \ ,
\label{power}
\end{equation}
corresponding to the binary correlator 
\begin{eqnarray}
\chi(l) = \left( \sqrt{2} + 1 \right) \frac{\sqrt{2} + (-1)^{l+1}}{1 - 16 l^{2}} 
\end{eqnarray}
which, due to the power law decay, represents long-range correlated disorder. The low wave number form of this power spectrum is $W\sim \mu$, therefore, $\beta=1$ and the asymptotic thermal conductivity must be independent of the system size for fixed boundary conditions, whereas for free boundary conditions $\kappa\sim N^{2/3}$. These analytical predictions are corroborated in Figs.~\ref{figuFC} and~\ref{figuLC}, where the results are still valid even for strong coupling between Langevin heat baths and the harmonic chain. Notice also that for both boundary conditions, the asymptotic thermal conductivity is practically the same for both models of heat baths when $\gamma=1$. However, as the coupling $\gamma$ becomes stronger, the asymptotic thermal conductivity is reached only for larger system sizes $N$, this phenomenon can also be observed from Eq.~(\ref{nosen}).  

It is important to stress that the low wave number behavior of the power spectrum provides information of the large scale decay of the correlation function. Indeed, if the power spectrum has a power law singularity in the low frequency regime $W(\mu)\sim \mu^{\beta}$ for $-1<\beta<0$, this corresponds to long range correlated disorder, where the binary correlator $\chi(l)\sim |l|^{-1-\beta}$.  Harmonic chain models with L\'evy-type disorder~\cite{ZLW15,HM19} and a harmonic chain whose  masses  follow a random sequence describing the trace of a fractional Brownian motion~\cite{MCRL03} are examples of systems with the aforementioned low wave number form of the power spectrum. Nevertheless, there exist correlated disorder models for which the low wave number behavior of the  power spectrum does not behave as a power law and, therefore, a non-standard thermal conductivity may arise as it is discussed below.


\subsection{Emergence of the non-standard thermal conductivity}

\begin{figure}[!t]
\centering
\includegraphics[width=\columnwidth]{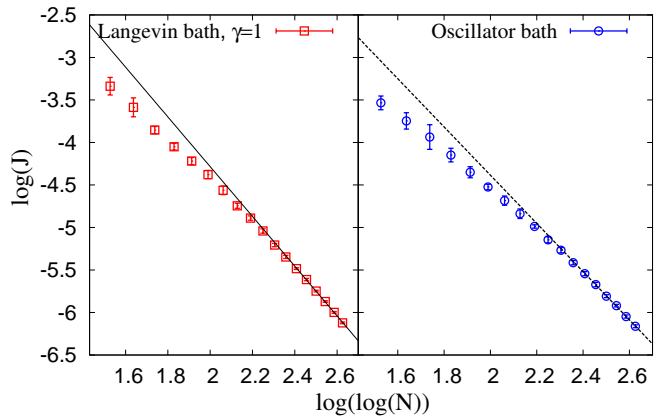}
\caption{(Color online) Natural logarithm of the heat flux as a function of the natural logarithm of the natural logarithm of the chain size for fixed boundary conditions. Different point styles represent the numerical data, while the continuous and dashed lines are the best fits of the numerical data, in the large size limit, with the function $f(N)=a\log(\log(N))+b$. For Langevin heat baths $a=-2.9\pm 0.03$ and $b=1.57 \pm 0.08$, while for oscillator heat baths $a=-2.88 \pm 0.03$ and $b=1.29\pm 0.05$.}
\label{Fig2F}
\end{figure}

\begin{figure}[!h]
\centering
\includegraphics[width=\columnwidth]{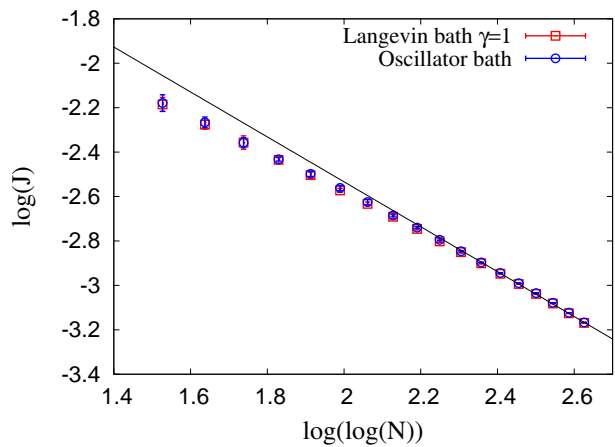}
\caption{(Color online) Natural logarithm of the heat flux as a function of the natural logarithm of the natural logarithm of the chain size for free boundary conditions. Different point styles represent the numerical data, while the continuous line is the best fits of the numerical data, in the large size limit, with the function $f(N)=a\log(\log(N))+b$. For both heat baths $a=-1.01\pm 0.01$ and $b=-0.51\pm 0.03$.}
\label{Fig2L}
\end{figure}

When the low frequency of the power spectrum doest not scale as a power law, the thermal conductivity may not be standard. An example of the latter statement is when the power spectrum has the form
\begin{equation}
W(\mu)=\frac{\pi A}{2}\exp\left(\frac{2}{\pi}A\right)\frac{1}{\mu^2}\exp\left(-\frac{A}{\mu}\right) ,
\end{equation}
being $A$ a parameter introduced in order to tune the numerical results, as will be seen below. Then if one uses  Eq.~(\ref{apro1}), the cut-off frequency acquires the following form
\begin{equation}
\omega_c=\frac{A\omega_{\text{max}}}{2(\log(N)-\log(C))},
\end{equation} 
where $C$ is a constant whose value is given by
\begin{equation}
C=\frac{16M^2\exp\left(-\frac{2}{\pi}A \right)}{\text{var}[m_n]\pi A}.
\label{cons}
\end{equation}
Therefore, if $\log(N)\gg 1$, $\omega_c$ scales as
\begin{equation}
\omega_c\sim \frac{1}{\log(N)},
\end{equation}
which leads to the following asymptotic form of the thermal conductivity:
\begin{eqnarray}
 \kappa\sim 
         \left\{
                \begin{array}{ll}
                {\displaystyle \frac{N}{\log(N)} }   & \mbox{for free BC} \vspace{0.2cm} \\     
                {\displaystyle \frac{N}{(\log(N))^3} } & \mbox{for fixed BC} 
                \end{array}
         \right. .
\label{final4}
\end{eqnarray} 
Now in general this asymptotic result is only reached for huge system sizes, but in order to observe this behavior in our numerical simulations we choose the parameter $A$ in such a way that the constant $C$ takes the value $C\approx1$. In this way, the asymptotic result~(\ref{final4}) can be observed for moderate system sizes ($N\sim 10^5$) as Figs.~\ref{Fig2F} and~\ref{Fig2L} show. There, $M=1$, $\gamma=1$, $k=1$, $\text{var}[m_n]=0.1$, $A=4$, $T_L=2$ and $T_R=1$. For fixed BC, $k_1=1$ for Model I and $k_1=0$ for Model II; whereas for free BC, $k_1=0$ for Model I and $k_1=1$ for Model II. To simulate the heat conduction in the harmonic chain we use formula~(\ref{flux}) which is an exact result for the harmonic chain model.

To create correlated disordered harmonic chains with the power spectrum given in Eq.~(\ref{final}) (or Eq.~(\ref{final4})), we use the standard technique of creating discrete colored noise from discrete white noise where the former is obtained as a linear combination of the latter
\begin{equation}
m_n=\sum^{\infty}_{n'=-\infty} G(n')x_{n'+n}+M,
\end{equation} 
with $\overline{x_n}=0$, and $\overline{x_nx_{n'}}=\delta_{n,n'}$. The modulation function $G(n)$ is given in terms of the pre-defined power spectrum
\begin{equation}
G(n)=\frac{2}{\pi}\int^{\pi/2}_0\sqrt{\text{var}[m_n]W(2\mu)}\cos(2\mu n).
\end{equation}


\section{Conclusion}

We study the localization and heat conduction properties of a harmonic chain with correlated isotopic disorder attached at its ends to Langevin or oscillator heat baths. When the power spectrum of the fluctuations of the random masses $W(\mu)$ scales in the low waver number limit as a power law, the asymptotic thermal conductivity $\kappa$ scales with the system size $N$ as $\kappa\sim N^{\alpha}$, which is a standard results for one dimensional systems. Here the exponent $\alpha$ depends on the boundary conditions, but not on the heat baths analyzed in this work. When $W(\mu)$ does not show a power law behavior in the low wave number limit, the thermal conductivity may not be given as a power law function of the system size. To corroborate the latter statement, the heat conduction properties of harmonic chains with a power spectrum of the form $W(\mu)\sim 1/\mu^2\exp(-1/\mu)$ were analyzed. The latter power spectrum yields to non-standard forms of thermal conductivity:  $\kappa \sim N/(\log N)^3$ for fixed boundary conditions and $\kappa \sim N/\log(N)$ for free boundary conditions. Our results for the asymptotic thermal conductivity do not depend neither on the spectral properties of the thermal baths analyzed in this work nor on the impedance mismatch between the baths and the harmonic chain. Thus, the results derived recently in Ref.~\cite{HM19} for the asymptotic thermal conductivity in harmonic chains with L\'evy-type disorder attached at its ends to Langevin heat baths are also valid for oscillator baths. In addition, we validated our analytical estimations using numerical simulations.

The importance of using generalized Langevin equations to obtain analytical results opens the possibility to study the effects of correlated mass disorder in the quantum regime, since Eq~ (\ref{flux}) is the classical limit of a more general equation~\cite{D08}. Furthermore, in Ref.~\cite{AABOI19} a dimensionless scaling parameter that depends on the temperature scale and localization length, for which the thermal conductance follows a universal behavior, was found. Therefore, this opens the question of whether a universal behavior exists for the thermal conductivity for the correlated mass disorder too and if a unified description exists for both mass and spring disorder.

\begin{acknowledgments}
J.A.M.-B. thanks support from 
FAPESP (Grant No.~2019/06931-2), Brazil, and
VIEP-BUAP (Grant No.~100405811-VIEP2019) and
PRODEP-SEP (Grant No.~511-6/2019.-11821), Mexico.
I.F.H.-G. thanks Prof. L. Tessieri for useful discussion and comments.
\end{acknowledgments}

\end{document}